\documentclass[prd,letterpaper,twocolumn,preprintnumbers,nofootinbib]{revtex4}

\usepackage{slashed}
\usepackage{amsmath,amssymb}
\usepackage{graphicx}
\usepackage{units}
\usepackage{xcolor}
\usepackage[hyperfootnotes=false,colorlinks,citecolor=blue]{hyperref}

\usepackage{comment}
\usepackage{ulem}

\newcommand{\beq}{\begin{equation}}
\newcommand{\eeq}{\end{equation}}
\newcommand{\bea}{\begin{eqnarray}}
\newcommand{\ena}{\end{eqnarray}}

\def \epsilon {\varepsilon}

\allowdisplaybreaks

\begin{document}

\title{
$W$ boson mass shift, dark matter and $(g-2)_\ell$ in a scotogenic-Zee model
}

\author{\bf Ritu Dcruz}
\email[E-mail: ]{rdcruz@okstate.edu}
\affiliation{Department of Physics, Oklahoma State University, Stillwater, OK 74078, USA}

\author{\bf Anil Thapa}
\email[E-mail: ]{wtd8kz@virginia.edu}
\affiliation{Department of Physics, University of Virginia,
Charlottesville, Virginia 22904-4714, USA}


\begin{abstract}

We present a singly charged scalar extension of the Scotogenic model, ScotoZee, which resolves the recently reported deviations in $W$ boson mass as well as lepton $g-2$. The model admits a scalar or a fermionic dark matter while realizing naturally small radiative neutrino masses. The mass splitting of $\sim 100$ GeV, required by the shift in $W$ boson mass, among the inert doublets fields can be evaded by its mixing with the singlet scalar, which is also key to resolving $(g-2)_\ell$ anomaly within $1\sigma$. We establish the consistency of this framework with dark matter relic abundance while satisfying constraints from charged lepton flavor violation, direct detection as well as collider bounds. The model gives predictions for the lepton flavor violating $\tau\to\ell\gamma$ processes testable in upcoming experiments.

\end{abstract}

\maketitle


\section{Introduction}
The CDF collaboration at Fermilab~\cite{CDF:2022hxs} reported a precision measurement of $W$ boson mass,
$M_W^\textrm{CDF}= (80.4335 \pm 0.0094) \;\textrm{GeV}$, which is in tension with the Standard Model~(SM) prediction, $M_W^\textrm{SM}= (80.357 \pm 0.004)\;\textrm{GeV}$ \cite{Awramik:2003rn}, with an excess at $7\sigma$ level, which may be an indication of new physics~(NP) beyond the Standard Model~(SM). The new result from CDF collaboration, with a much reduced uncertainty, has a higher precision than the PDG world average of $M_W^\textrm{PDG}= (80.377 \pm 0.012)\;\textrm{GeV}$~\cite{ParticleDataGroup:2022pth} which takes into account the $W$ mass measurements from LEP~\cite{ALEPH:2013dgf}, Tevatron~(CDF~\cite{CDF:2012gpf} and D0~\cite{D0:2012kms})~\cite{CDF:2013dpa} and LHCb collaboration~\cite{LHCb:2021abm}. The PDG average, which is in agreement with the SM prediction, disagrees with the CDF Run-II result. It has been shown that the improvement in parton density functions \cite{Gao:2022wxk} and perturbative matrix elements \cite{Martin:2022qiv,Isaacson:2022rts} cannot account for this discrepancy. However, the discrepancy may be due to high-twist power corrections within the SM that are normally not considered in perturbative calculations \cite{ELLIS19821,Ellis:1982cd}. This work proceeds under the assumption that the new CDF measurement will be validated as the right result of $W$ boson mass.

Some possible explanations to the $W$ boson mass shift  can arise at tree-level~\cite{Strumia:2022qkt,Asadi:2022xiy,Bagnaschi:2022whn,DiLuzio:2022ziu,Chen:2022ocr,PerezFileviez:2022gmy,Ghoshal:2022vzo,Borah:2022zim,Athron:2022isz,Almeida:2022lcs,Addazi:2022fbj,Heeck:2022fvl,Du:2022fqv,Zeng:2022lkk,Cheng:2022aau,Cai:2022cti}, or at loop level~\cite{Liu:2022jdq,Song:2022xts,Biekotter:2022abc,Crivellin:2022fdf,Heo:2022dey,Bahl:2022xzi,Ahn:2022xeq,Carpenter:2022oyg,Popov:2022ldh,Ghorbani:2022vtv,Lu:2022bgw,Han:2022juu,Heckman:2022the,Cao:2022mif,Lee:2022gyf,Abouabid:2022lpg,Benbrik:2022dja}, along with the prospect of reconciling one or more discrepancies \cite{Baek:2022agi,Bhaskar:2022vgk,Chowdhury:2022moc,Cheung:2022zsb,Babu:2022pdn,Lee:2022nqz,Athron:2022qpo,Yang:2022gvz,Du:2022pbp,Zhang:2022nnh,Zhu:2022scj,Cacciapaglia:2022xih,Fan:2022dck,Zhu:2022tpr,Batra:2022org,Arcadi:2022dmt,Nagao:2022oin,Kawamura:2022uft,Tang:2022pxh,Zhou:2022cql,Botella:2022rte,Kim:2022xuo,Kim:2022hvh} such as flavor anomalies and dark matter. Several other papers~\cite{deBlas:2022hdk,Campagnari:2022vzx,Blennow:2022yfm,Sakurai:2022hwh,Fan:2022yly,Arias-Aragon:2022ats,Paul:2022dds,Gu:2022htv,DiLuzio:2022xns,Cheng:2022jyi,Endo:2022kiw,Du:2022brr,Balkin:2022glu,Krasnikov:2022xsi,Zheng:2022irz,Sun:2022zbq,Peli:2022ybi,Kanemura:2022ahw,Mondal:2022xdy,Tan:2022bip,Cirigliano:2022qdm,Borah:2022obi,Yang:2022qgs,Rahaman:2022dwp,Pellen:2022fom,Dermisek:2022xal,Chen:2022ntw,Perez:2022uil,Gupta:2022lrt,Isaacson:2022rts,Yuan:2022cpw} also examine the consequence of the CDF $M_W$ anomaly on new physics scenarios.

Independently, muon $(g-2)$ collaboration at Fermilab~\cite{Abi:2021gix} has confirmed the long standing discrepancy in the anomalous magnetic moment (AMM) of muon measurement at BNL in 2006 \cite{Bennett:2006fi} at a combined $4.2\sigma$\footnote{The recent results from the BMW collaboration~\cite{Borsanyi:2020mff} agrees with the experimental measurement within $1.6\sigma$.} deviation, $\Delta a_\mu^\textrm{exp} = (2.51\pm 0.59)\times 10^{-9}$, from the SM prediction (see Ref.~\cite{Aoyama:2020ynm} and the references therein). In addition to these recent anomalies, astrophysical and cosmological observations~\cite{Planck:2013pxb,WMAP:2010qai,Planck:2018vyg} present a compelling evidence for the existence of dark matter (DM), for  which the SM fails to provide an explanation. Moreover, one of the major shortcomings of the SM is its inability to explain the origin of non-zero neutrino mass substantiated by several experiments~\cite{ParticleDataGroup:2020ssz}.

In this work we show that  a simple extension of the Scotogenic model ~\cite{Ma:2006km} with a charged singlet (ScotoZee model) can simultaneously address all the puzzles previously mentioned. Our novel ScotoZee model\footnote{The scalar content is the same as the Inert Zee model~\cite{Longas:2015cnr,Longas:2015sxk} with only right-handed neutrinos in contrast to vector-like singlets and doublets. The ScotoSinglet model~\cite{Beniwal:2020hjc} is a neutral scalar extension of the Scotogenic model. Neither models can resolve the discrepancy in $(g-2)_\mu$~\cite{Gaviria:2018cwb}.} is the simplest model that furnish a direct link between neutrino mass generation, dark matter, AMM of muon and also provide an upward mass shift in $W$ boson in agreement with the CDF measurement. Additionally, the  presumed anomaly in the AMM of electron \cite{Hanneke:2008tm,Parker:2018vye,Aoyama:2017uqe} can also be addressed within the same framework. 
We explore the parameter space of the ScotoZee model spanned by both the bosonic and fermionic DM candidates, while being consistent with the current experimental constraints.

The rest of the paper is organized as follows. In Sec.~\ref{sec:model} we give a brief description of the ScotoZee model, discussing the neutrino mass generation and the scalar sector. In Sec.~\ref{sec:Wmass} and Sec.~\ref{sec:g2} we respectively introduce and examine a resolution to $W$ mass shift and AMM phenomenology in the model. Sec.~\ref{sec:DM} discusses DM phenomenology for both scalar and fermionic DM candidates. Lastly, we integrate the three puzzles ($W$-mass shift, lepton $g-2$, and DM) with neutrino mass generation and LFV constraints invoking a highly predictive flavor texture in Sec.~\ref{sec:nut} before concluding in Sec.~\ref{sec:conclusion}. 

\section{Model}\label{sec:model}
The proposed ScotoZee model is  a simple charged singlet $S^+~(1,1;-)$ extension of the Scotogenic~\cite{Ma:2006km} model, which contains Majorana singlet fermions $N_{R_i}~(1,0;-)$ and the scalar doublet $(\eta^+, \eta^0)\equiv\eta~(2,1/2;-)$,
under the gauge group $SU(2)_L \times U(1)_Y \times \mathbb{Z}_2$. All the new particles are odd under $\mathbb{Z}_2$ while the SM particles are even, guaranteeing the stability of the DM candidate; the lightest among the new neutral $\mathbb{Z}_2$-odd particles. The charged scalar singlet $S^+$ not only gives corrections to anomalous magnetic moment of muon and electron through the mixing with charged doublet, but also serves as a portal to generate correct relic abundance for fermionic DM.

\begin{figure}[!t]
        \includegraphics[width=0.35\textwidth]{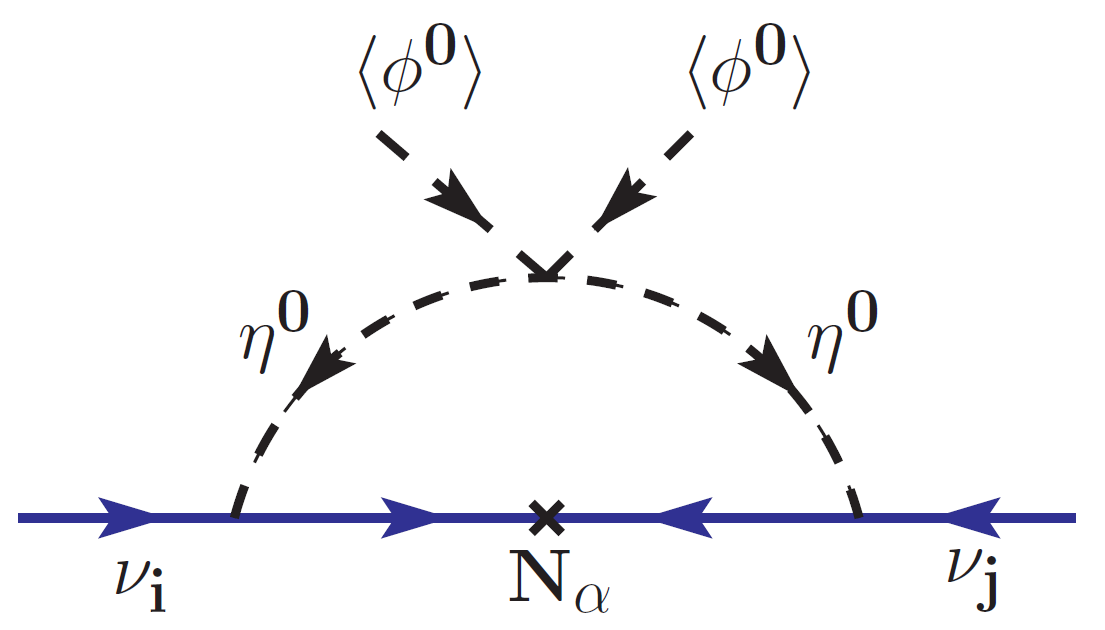}\\
    \includegraphics[width=0.35\textwidth]{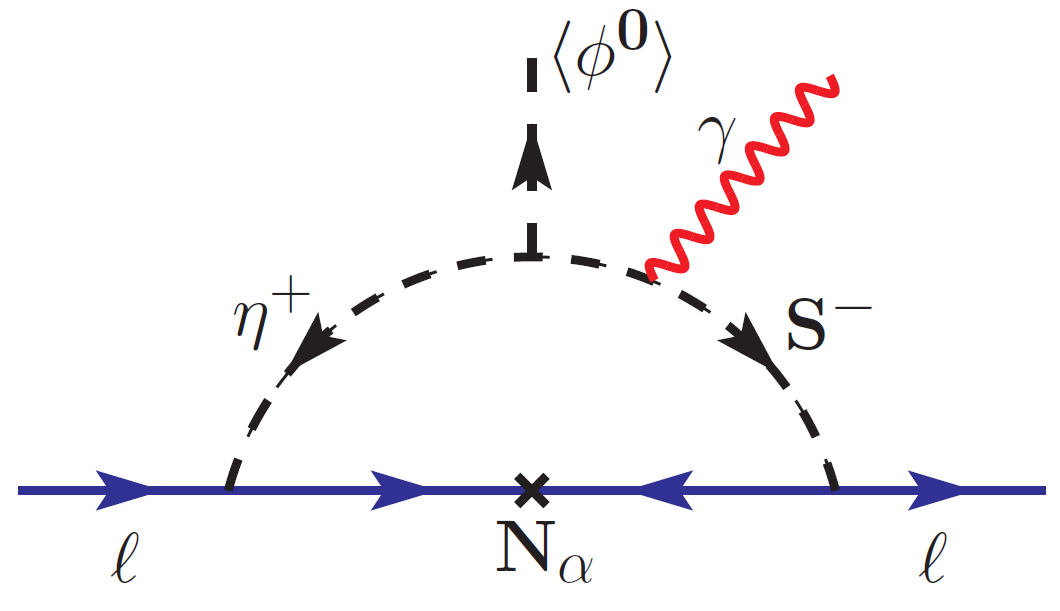}
    \caption{Radiative neutrino mass generation at one-loop (top). The dominant correction to AMMs arising through the chiral enhancement (bottom). The cross ({\boldmath $\times$}) represents mass insertion whereas $\ell=e~(\mu)$ for electron (muon) AMM. }
    \label{fig:feynman}
\end{figure}

The effective Yukawa Lagrangian in the extended model can be written as
        \begin{equation}
  -\mathcal{L}_Y\supset    Y_{ij}\overline{L}_{L_i}\widetilde{\eta}N_{R_j}+f_{ij}\overline{\ell}_{R_i}S^-\overline{N}_{R_j}
             +\text{h.c.}.\label{eq:Lyukawa}
        \end{equation}      
The $\mathbb{Z}_2$ symmetry, being exact, prevents $\eta^0$ from obtaining a non-zero vacuum expectation value (VEV) and neutrinos remain massless at tree-level. Moreover, the SM Higgs $h$ is decoupled from the new $\mathcal{CP}$-even ($Re(\eta^0)\approx H$) and -odd ($Im(\eta^0)\approx A$) scalars. The tree-level scalar potential of the model is given by
\begin{align}
            V&=\mu_h^2 \phi^\dagger \phi +\mu_S^2 S^-S^+ + \mu_\eta^2\eta^\dagger\eta+\frac{\lambda_1}{2}(\phi^\dagger\phi)^2+\frac{\lambda_2}{2}(\eta^\dagger\eta)^2 \nonumber\\
            &
            +\lambda_3(\phi^\dagger\phi)(\eta^\dagger\eta)+\lambda_4(\phi^\dagger\eta)(\eta^\dagger\phi)+\frac{\lambda_5}{2}\{(\phi^\dagger\eta)^2+\text{h.c.}\} \nonumber\\
            &+\frac{\lambda_6}{2}(S^-S^+)^2 +\lambda_7(\phi^\dagger\phi)(S^-S^+)+\lambda_8(\eta^\dagger\eta)(S^-S^+)\nonumber\\
            &
            +\frac{\mu}{2}\{\epsilon_{\alpha\beta}\phi^\alpha\eta^\beta S^-+\text{h.c.}\}.
            \label{eq:potential}
\end{align}
The charged scalars $\{\eta^+,S^+\}$ mix  giving rise to mass eigenstates $\{H_1^+,H_2^+\}$. The masses of  the scalar fields in the physical basis are given by
\begin{align}
    m^2_h=\lambda_1 v^2,\, \hspace{4mm} m^2_{H(A)}=\mu_{\eta}^2+\frac{v^2}{2}\left(\lambda_3+\lambda_4\pm \lambda_5\right),\nonumber\\
   m_{H^+_i}^2=   \frac{1}{2}\left(\mu_2+\mu_3\pm \sqrt{(\mu_2-\mu_3)^2+2\mu^2v^2}\right),
\end{align}
where $\mu_2=\mu_\eta^2+\frac{\lambda_3}{2}v^2$, $\mu_3=\mu_S^2+\frac{\lambda_7}{2}v^2$. Here $\mu_{\eta, S}$, $\lambda_i$, and $\mu$ are the bare-mass terms, quartic couplings, and cubic coupling, respectively. The mixing angle between the charged scalar fields is defined as
\begin{equation}
    \sin2\theta=\frac{-\sqrt{2}\mu v}{m_{H_1^+}^2-m_{H_2^+}^2} \, ,
\end{equation}
with the VEV, $v \simeq 246$ GeV.
In this work, we comply with the perturbative and vacuum stability conditions~\cite{Deshpande:1977rw,Kannike:2012pe} constraining the scalar couplings. We also have ensured that our parameter space does not drive the the mass parameters $\mu_\eta^2$ and $\mu_S^2$ to negative values; negative bare mass terms can lead the inert doublet or the charged singlet to attain a non-zero VEV, thereby breaking the $\mathbb{Z}_2$ symmetry ~\cite{Merle:2015gea}. Such $\mathbb{Z}_2$ parity violation not only destabilizes the DM candidates but will also allow unacceptably large tree-level neutrino masses. These arise from the RG evolution of the bare mass terms and might pose a serious problem when the coupling of the inert scalars to heavy neutrinos are of $\mathcal{O}(1)$. As will be discussed in the following sections, the coupling of $\eta$ to $N_{R_k}$ are small enough to avoid breaking the $\mathbb{Z}_2$ symmetry. On the other hand, the coupling between $S^+$ and $N_{R_k}$ are taken to be $\mathcal{O}(1)$. Thus, in order for the model to be valid above the EW scale (such that the $\mathbb{Z}_2$ symmetry is preserved), it is important to check that the RG evolution does not only preserve $\mu_\eta^2 > 0$ but also $\mu_S^2 > 0$. The trilinear coupling $\mu$ gives positive contributions to the RG evolution of the new scalar mass-squared parameters, thereby helping prevent the breaking of the model symmetries. Note that
it is also essential to ensure that $\mu$ is not much larger than the scalar masses as it can result in a deeper minimum than the SM one~\cite{Barroso:2005hc,Babu:2019mfe}.
The Majorana mass term $\frac{1}{2}M_{N_i} N_i N_i$ along with the scalar quartic term $\frac{\lambda_5}{2}\{(\phi^\dagger\eta)^2+\text{h.c.}\}$ breaks the lepton number by two units generating one-loop neutrino mass~(c.f. Fig.~\ref{fig:feynman}~(top)) $\mathcal{M}_\nu$  expressed as
\begin{align}
    &\left(\mathcal{M}_\nu\right)_{ij}= \sum_k Y_{ik}\Lambda_{k}Y^*_{k j} \nonumber \, ,\\
    &\Lambda_k = \frac{M_{N_k}}{16\pi^2}\Bigg[\frac{m_H^2}{m_H^2-M_{N_k}^2} \log\frac{m_H^2}{M_{N_k}^2} - (m_H \leftrightarrow m_A)\Bigg]. 
    \label{eq:numass}
\end{align}

Here the lightest mass eigenstates $\{ H, A\}$ and $N_i$ can serve as a viable bosonic and fermionic DM candidates, respectively. It is important to point out that unlike the Scotogenic model, where $M_{N}$ can be at canonical seesaw scale of $10^9$ GeV or the Yukawa coupling $Y$ arbitrarily small, the $(g-2)_\mu$ in the model requires the scale to be in the (sub) TeV range along with $\mathcal{O}(0.1-1.0)$ Yukawa coupling. Thus, a successful explanation of $m_\nu\sim 0.1$ eV would naturally require $m_H$ to be nearly degenerate with $m_A$.


\section{Correction to $W$~boson mass}\label{sec:Wmass}
The shift in $W$ boson mass~\cite{Grimus:2008nb} can be evaluated as a function of the oblique parameters, $S,\, T$ and $U$ ~\cite{Peskin:1990zt,Peskin:1991sw} that quantify the deviation of a new physics model from the SM through radiative corrections arising from shifts in gauge boson self energies.
The oblique parameters in our model get corrections from the extended Higgs sector which is same as in the Zee model~\cite{Zee:1980ai} except for the $\mathbb{Z}_2$ charge preventing the mixing with the SM Higgs doublet. Therefore, we use the expressions for $S$, $T$ and $U$ given in ~Ref.~\cite{Herrero-Garcia:2017xdu} under the alignment limit~\cite{Haber:2010bw}. Note that the corrections to $U$ at one-loop level is suppressed compared to $S$ and $T$. 

\begin{figure}[!t]
    \includegraphics[width=0.45\textwidth]{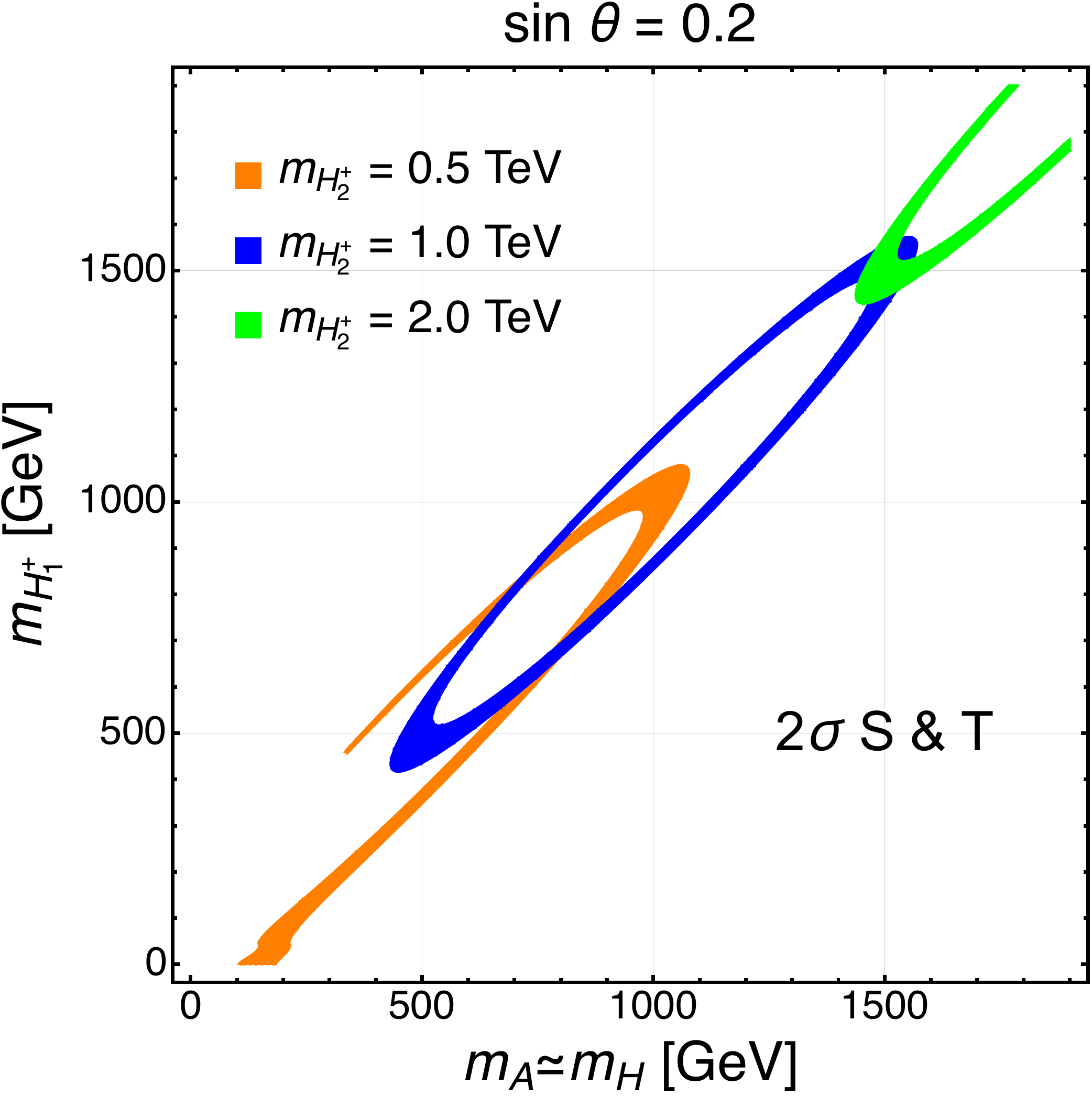}\\ 
    \vspace{4mm}
    \includegraphics[width=0.45\textwidth]{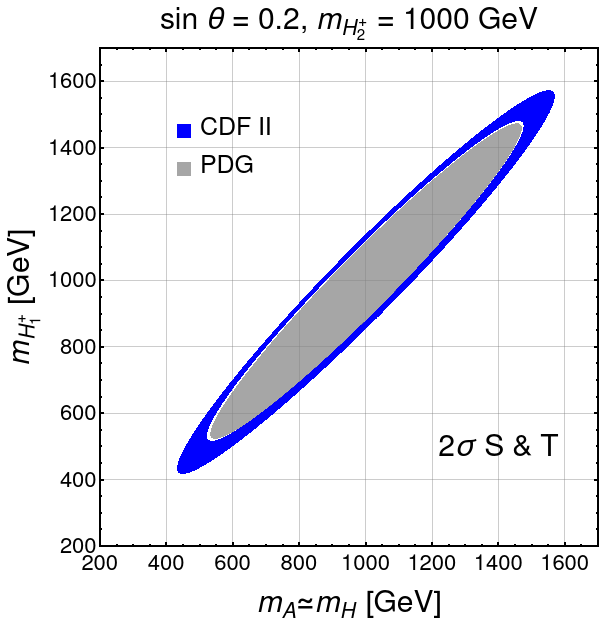}
    \caption{Top figure shows the mass splitting between the components of doublet scalar required by the new CDF measurement of $W$ boson mass using $2\sigma$ ranges for $S$ and $T$ from Ref.~\cite{Asadi:2022xiy}. Bottom figure shows the $2\sigma$ band allowed by the CDF measurement in contrast with the PDG world average~\cite{ParticleDataGroup:2022pth,Lu:2022bgw}.}
    \label{fig:splittings}
\end{figure}

With the new precision measurement of $M_W$ by CDF, some electroweak~(EW) observables are expected to suffer from this deviation. 
We incorporate the global EW fit~\cite{Asadi:2022xiy} with the new CDF data to quote the $2\sigma$ allowed ranges of oblique parameters. We confirm the necessity of mass splitting in 2HDM~\cite{Ahn:2022xeq,Bahl:2022xzi} to accommodate the recent CDF results and show that the introduction of the charged singlet scalar allows the components of the doublet field to be degenerate in certain regions for a specific choice of $\sin\theta$ and $m_{H_2^+}$, as can be seen from Fig.~\ref{fig:splittings} (top) (for more detail see Fig.~\ref{fig:Tparam2}). 
The splitting $\delta_{H+}=m_{H_1^+}-m_H$ depends on the mixing angle, for instance, it can be at most $\sim 140$ GeV for $\sin \theta = 0.2$. In spite of explaining the CDF $W$ mass shift, the proposed model remains consistent with the previous experiments resulting in the PDG world average of $W$ boson mass. It can be seen from Fig~\ref{fig:splittings} (bottom) that region consistent with the older experiments opens up more parameter space of the model, even allowing for degeneracy between the doublet fields in the whole range of allowed parameter space for particular values of $\sin\theta$ and $m_{H_2^+}$. 

\section{Anomalous Magnetic Moment}\label{sec:g2}
The charged scalar contributions to anomalous magnetic moment at one-loop \cite{Leveille:1977rc} as shown in Fig.~\ref{fig:feynman}~(bottom) is 
\begin{align}
        \Delta a_{\ell}^{H^+_{1}}=&\frac{m_{\ell}^2}{16\pi^2}\Big(\left(|Y_{\ell i}|^2\sin^2\theta+|f_{\ell i}|^2\cos^2\theta\right)\ G[m_{H_1^+},2] \nonumber\\
        &+ \frac{M_{N_i}}{m_\ell}\ \textrm{Re}(Y_{\ell i}f^*_{\ell i})\sin2\theta\ G [m_{H^+_1},1]\Big),
\end{align}
where,
\begin{equation}
    \begin{aligned}
    G[M,\epsilon]=\int_{0}^{1} \frac{x^\epsilon(x-1)\ \,dx }{m_\ell^2x^2+(M^2-m_\ell^2)x+M_{N_i}^2(1-x)}, 
    \end{aligned}
\end{equation} 
and  $\Delta a_{\ell}^{H^+_{2}}=\Delta a_{\ell}^{H^+_{1}}(\theta \to \frac{\pi}{2}+\theta)$.
The dominant contribution to $\Delta a_\mu$ comes from the Majorana neutrino mass enhancement aided by the mixing of the charged scalar mediators as shown in Fig.~\ref{fig:feynman}~(bottom). The sign of the product of Yukawa couplings and the mixing angle can be chosen independently. This in turn allows for the simultaneous explanations of $\Delta a_\ell$ ($\ell=e,\, \mu$). Moreover, $\Delta a_\mu$ provides an upper limit on the mass of Majorana neutrino (charged scalar) of order 15 (6.5) TeV with $f, Y\sim \mathcal{O}(1)$. The mass limit is relaxed in the case of $\Delta a_e$. 

Note that the Yukawa couplings and the masses of charged scalars are severely restricted by the charged lepton flavor violating~(cLFV) processes such as radiative decay $\ell_i \to \ell_j \gamma$ \cite{Lavoura:2003xp}; such processes are enhanced in our model by the mass insertion of Majorana neutrinos. Moreover, although trilepton decays such as $\mu\to 3e$ do not occur at the tree-level, they arise at the one-loop with large branching ratios~\cite{Toma:2013zsa}. The same is also true for $\mu-e$ conversion in the nuclei. We impose these constraints in our parameter scan.


\section{DM Phenomenology}\label{sec:DM}
\begin{figure}[t]
    \includegraphics[width=0.49\textwidth]{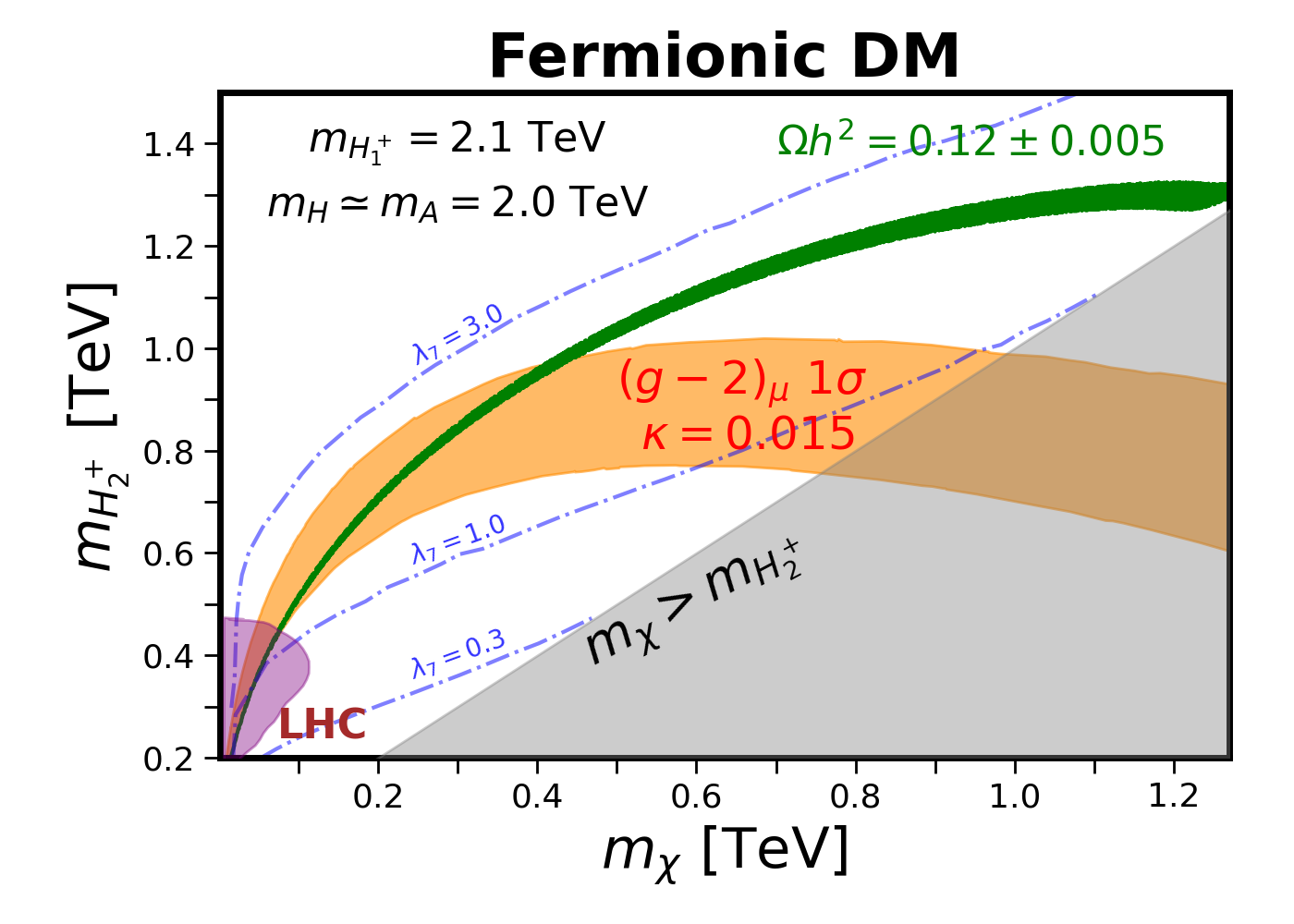}
    \includegraphics[width=0.49\textwidth]{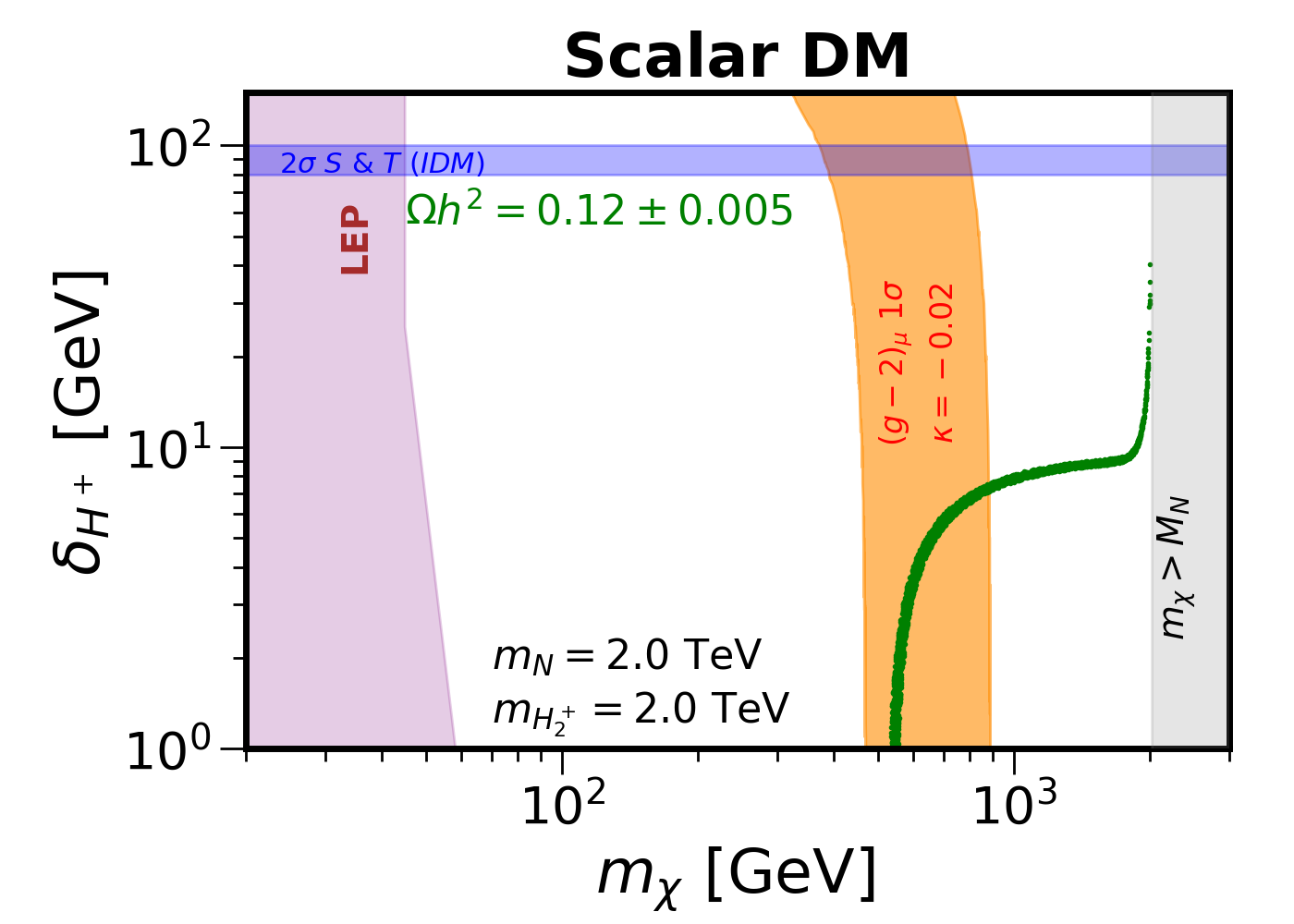}
    \caption{The allowed parameter space for fermion~(top) and neutral scalar~(bottom) dark matter candidate to simultaneously explain the AMM of muon and $W$ boson mass shift. The green (orange) band satisfies the correct relic abundance ($(g-2)_\mu$) for fixed ~$\kappa=Y^*f\sin\theta$. The choice of scalar masses provide an upward $W$ boson mass shift as expected from CDF measurement. The purple band (top figure) is the exclusion region region from $pp \to \ell^+ \ell^- + \slashed{E}_T$ signature obtained by recasting supersymmetric slepton searches at the LHC \cite{CMS:2018eqb} and from chargino production searches at LEP~\cite{OPAL:2003wxm,OPAL:2003nhx}
    ~(bottom figure). The blue dashed-dotted lines~(top) shows the direct detection cross section lower bounds~\cite{Hisano:2015bma,Herrero-Garcia:2018koq} from the LZ experiment~\cite{LZ:2022ufs} for different quartic coulings $\lambda_7$ and our choice of Yukawa coupling $f=1.5$.}
    \label{fig:moneyplot1}
\end{figure}
In addition to explaining $W$ boson mass shift and $\Delta a_\ell$, the proposed model can easily accommodate both the scalar (lightest of $H$ and $A$) and fermionic (lightest among $N_i$) dark matter candidates ($\chi$). We consider both scenarios and analyze the parameter space by implementing the model in \texttt{SARAH}~\cite{Staub:2008uz} and numerically evaluating the relic abundance using the software \texttt{MicrOMEGAs}~\cite{Belanger:2006is}. The relic density of DM is achieved through standard thermal freeze-out mechanism.

For the case of Majorana fermion as a DM ($\chi\equiv N$) candidate, the annihilation channel which determines the observed relic abundance are DM self-(co-)annihilation into charged leptons $\ell_\alpha^- \ell_\beta^+$ (light neutrinos $\nu_\alpha \bar{\nu}_\beta$) through $t$-channel processes mediated by the $\mathbb{Z}_2$-odd scalars, $H_i^+$ ($H,A$) via Yukawa couplings $Y$ and/or $f$. The neutrino oscillation data determines the flavor structure of Y making it natural to select relatively small $Y$ and heavy doublet scalar $\eta$ $\sim \mathcal{O}\ ({\rm TeV})$ such that the LFV constraints are relaxed. Thus, we choose $f_{ii} \sim\mathcal{O}(1)\ (i=1,2)$ and degenerate $N_i$ to maximize the contribution to annihilation mode $\chi \chi \to \ell \ell$ via $S^+$; the allowed parameter space in the mass plane can be seen in Fig.~\ref{fig:moneyplot1}~(top) along with the region resolving muon AMM for a specific choice of $\kappa = Y^* f \sin\theta = 0.015$.

Note that in Fig.~\ref{fig:moneyplot1}, the effects of co-annihilation are not taken into account for fermionic DM. The co-annihilation between fermionic DM and the charged scalar $H_2^\pm$ becomes important when the mass difference is small; $(m_{H_2^\pm}-m_N)/m_N< 0.1$~\cite{Jueid:2020yfj}. This dominates over the annihilation processes mainly when $f<1$, a parameter space not favorable for our work to incorporate $(g-2)_\ell$ and satisfy LFV. Thus, we simply avoid co-annihilation by taking the mass splitting larger.

Although fermionic DM does not contribute to tree-level direct detection cross section, it can arise at one-loop order via photon, $Z$, and Higgs penguin diagrams involving charged scalars and charged leptons in the loop~\cite{Herrero-Garcia:2018koq}. The relevant dimension-6 operators for Majorana DM direct detection are the ones with bilinears $\overline{\chi}\chi$, $\overline{\chi}\gamma_5\chi$, and $\overline{\chi}\gamma^\mu\gamma_5\chi$. In the model presented here, both Yukawa couplings $f$ and $Y$ give rise to such diagrams. As previously mentioned the Yukawa coupling $Y$ needs to be small. Thus the dominant contribution comes from the coupling $f \sim \mathcal{O}(1)$ of Eq.~\eqref{eq:Lyukawa} with right-handed charged lepton $\ell_R$ and charged scalar $S^+$ in the loop. In this case, the $Z$ penguin contribution is suppressed due to the absence of axial-vector coupling to $\ell_R$. Photon mediated processes can lead to anapole operator~\cite{Ho:2012bg} of the form $\mathcal{O}^q_{AV}=(\overline{\chi}\gamma^\mu\gamma_5\chi)(\overline{q}\gamma_\mu q)$, where $q$ represent the quark flavor eigenstates. Anapole contributions, however, are momentum and velocity-suppressed in the non-relativistic limit~\cite{Gresham:2014vja}. We find that the cross section from anapole contribution for the DM masses in our framework is at most of $\mathcal{O}(10^{-55})\  \text{cm}^2$~\cite{Bai:2014osa,Herrero-Garcia:2018koq}, well beyond the current sensitivity.  The dominant contribution, therefore, appears from the Higgs penguin diagram leading to spin-independent direct detection from the operator $\mathcal{O}^q_{SS}=m_q(\overline{\chi}\chi)(\overline{q}q)$. The corresponding Wilson coefficient is dependent not only on the Yukawa coupling $f$ but also on the quartic coupling $\lambda_7$. This contribution can therefore be suppressed by choosing $\lambda_7$ small, allowing for $f$ coupling to be $\gg 1$. The allowed parameter space with the current bounds~\cite{LZ:2022ufs} for  direct detection cross sections~\cite{Hisano:2015bma,Herrero-Garcia:2018koq} is shown in Fig.~\ref{fig:moneyplot1}~(top) for different choices of $\lambda_7$ and a fixed Yukawa coupling $f=1.5$.

In the case of scalar dark matter, which we choose to be the $\mathcal{CP}$-even $H \equiv \chi$ (nearly degenerate\footnote{The mass splitting is of order $\mathcal{O} (100)$ keV~\cite{Arina:2009um} to evade direct detection.} with $A$ and $\lambda_5 < 0$), pair of DM can annihilate to $W^+ W^-$, $ZZ$, $\nu_\alpha \nu_\beta$, $hh$, $\bar{\ell} \ell$, and $\bar{q}q$. The low mass regime suffer a strong constraint form LEP \cite{OPAL:2003wxm,OPAL:2003nhx,Pierce:2007ut} 
which can be satisfied if one assumes $m_\chi > M_Z/2$, $m_{H_1^+}>M_W/2$ and $m_{H_1^+}+m_\chi > M_W$. For larger DM mass, it predominantly annihilates to a pair of $W^+W^-$ and $ZZ$, for which the allowed region is $m_\chi\gtrsim 500$ GeV and mass splitting $\delta_{H^+}=m_{H_1^+}-m_\chi \lesssim 30$ GeV as shown in Fig.~\ref{fig:moneyplot1} (bottom). This can be relaxed by making the Higgs quartic coupling $\gtrsim 1$, a choice strongly constrained by direct detection bound \cite{Lehnert:2018vzs,LUX-ZEPLIN:2018poe,XENON:2018voc,CDMS-II:2009ktb,LZ:2022ufs}. 

In this work we take the quartic couplings $\lambda_3+\lambda_4+\lambda_5 \ll 1$ to automatically satisfy direct detection bound obtained from the scalar DM interacting with nucleus at the tree-level through the SM Higgs boson.  Moreover, it is favored to take the couplings $Y_{ij}$ small and $M_N\sim\mathcal{O}$(TeV) to be consistent with neutrino fit, which implies that the DM analysis is indistinguishable from the known inert doublet model (IDM). It turns out that the CDF measurement requires mass splitting among the inert doublet fields of $\mathcal{O}(100)$ GeV, disfavouring the scalar DM candidates in the Scotogenic/IDM. However, the mixing between the charged scalars in this model allows the components of the doublet field to be degenerate~(c.f. Fig~\ref{fig:splittings}), thereby admitting the $\mathcal{CP}$-even $H$ to be a viable DM candidate, as shown in Fig.~\ref{fig:moneyplot1} (bottom). 

The direct detection cross section can arise at one-loop from vertices like $HHZZ$, $HHWW$, $HZA$, and $HWH_1^\pm$ driven by the gauge coupling. These contributions can be as important as the tree-level processes. In the model presented here, the quartic couplings that give rise to tree-level direct detection is chosen to be small, thus its contribution in comparison to loop effect is negligible. For the masses $m_\chi\geq 500$ GeV required to explain the total observed relic density within the model (see Figure 3 (bottom)), we find that the one-loop cross section is $\sim 1.1\times 10^{-46}$ cm$^2$ (see Ref.~\cite{Cirelli:2005uq,Klasen:2013btp,Banerjee:2019luv} for details). This evades the current experimental bound \cite{Lehnert:2018vzs,LUX-ZEPLIN:2018poe,XENON:2018voc,CDMS-II:2009ktb,LZ:2022ufs}, but will be sensitive to upcoming experiments~\cite{XENON:2020kmp,Church:2020env}. 


\section{Neutrino Fit/ Lepton Flavor violation}\label{sec:nut}
\begin{figure}[!t]
    \includegraphics[width=0.48\textwidth]{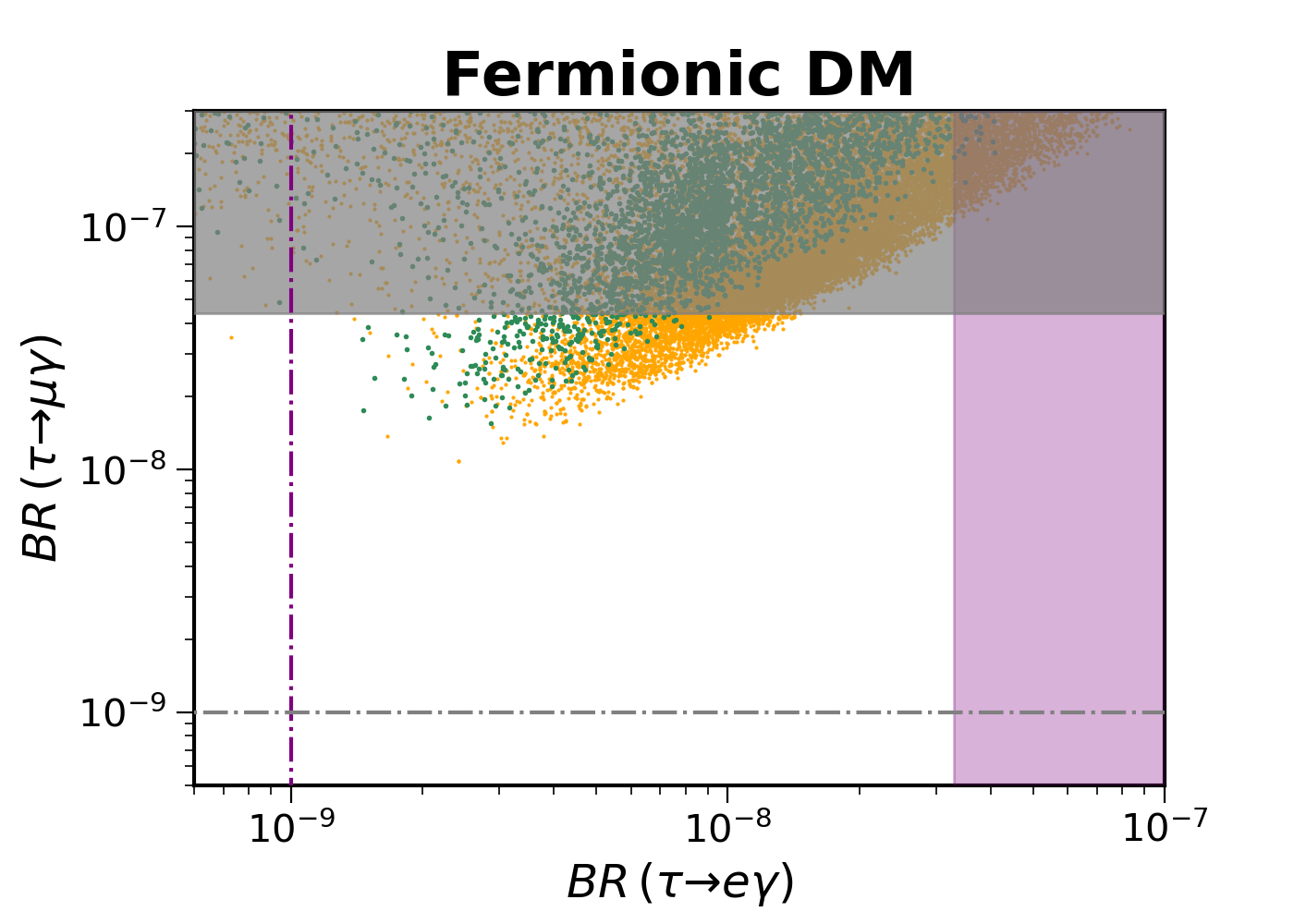}
    \caption{Scattered plot assuming the fermionic DM with the same parameter space given in Fig.~\ref{fig:moneyplot1} (top). Colored shaded regions are the current exclusion limits \cite{BaBar:2009hkt}, whereas dashed-dotted lines represent the future projected sensitivities \cite{Hayasaka:2013dsa}. The orange~(green) dots correspond to solutions that resolve $\Delta a_\mu$~($\Delta a_{e,\mu}$), satisfy the observed relic density as well as neutrino oscillation observables within their $2\sigma$ measured values \cite{Esteban:2020cvm}. }
    \label{fig:LFV}
\end{figure}
The neutrino mass formula of Eq.~\eqref{eq:numass}, lepton $g-2$ and the dark matter analysis have close-knit correlation through Yukawa couplings, Majorana fermions, and new scalars. As previously stated, $(g-2)_\ell$ sets an upper bound on the masses of the Majorana fermions and the charged scalars with $f,Y\sim\mathcal{O}(1)$. Moreover, the maximum splitting among the doublet fields is restricted by the shift in $W$-boson mass, thereby forcing the parameter space to the region $m_A \simeq m_H$, crucial in explaining the observed neutrino oscillation data.

In order to check the consistency with the neutrino oscillation data and efficiently probe the model with LFV observables, we adopt Casas-Ibarra parametrization~\cite{Casas:2001sr} to rewrite the Yukawa matrix $Y$ of Eq.~\eqref{eq:numass} in terms of neutrino mass parameters
\begin{equation}
    Y = \ U\sqrt{\mathcal{M}_\nu^{\rm diag}} R^\dagger \,\sqrt{\Lambda}^{-1}   \,,
\end{equation} 
where $R$ is an arbitrary complex orthogonal matrix. The neutrino oscillation parameters are scanned within the $2\sigma$ allowed ranges~\cite{Esteban:2020cvm} to obtain the Yukawa matrix. 

As mentioned earlier, the product of Yukawa couplings 
$Y_{\ell i}f_{\ell i}^*$ can explain $(g-2)_\ell$; however these couplings are constrained by LFV processes. To facilitate a direct correlation between the observed relic density and both muon and electron $(g-2)$, we consider two (degenerate) stable DM candidates with their respecting Yukawa couplings of $\mathcal{O}(1)$. This also allows more parameter space for the DM mass $m_\chi$ and mediator mass $m_{H_2^+}$ as shown in Fig~\ref{fig:moneyplot1}~(top). With such a choice of large $f_{ii} (i=1,2)$, the mass enhancement to $\ell_i \to \ell_j\gamma$ severely restricts the parameter space. Such chirally enhanced contribution can be suppressed with a suitable choice of Yukawa couplings and masses of the Majorana fermions. For instance, chirally enhanced $\mu\to e \gamma$ can be evaded with the choice of $Y_{12}=Y_{21}\sim0$ or $Y_{21}f_{11}\simeq -Y_{12}f_{22}$ for $M_{N_1}=M_{N_2} (=m_{\chi})$. We then check the consistency of our fit by computing the branching fractions of $\ell_i\to\ell_j\gamma$ and $\ell_i\to 3\ell_j$ process at one-loop level and make testable predictions for fermionic DM (see Fig.~\ref{fig:LFV}). In the case of scalar DM, since the Yukawa coupling $f$ does not play any role in relic abundance, there is more freedom in the choice of parameters and yield no sizeable predictions. It is important to note that a single Majorana fermionic DM candidate by itself can successfully resolve $(g-2)_\mu$ with $f_{\mu i}\sim\mathcal{O}(1)$. This would open up the parameter space with much weaker constraints arising from LFV processes. However, such a choice would not lead to direct correlation between $(g-2)_e$ and DM, and also does not lead to LFV prediction. \\ 

\section{Conclusions}\label{sec:conclusion}
In the light of recent experimental results confirming a $4.2\sigma$ discrepancy in the measurement of $(g-2)_\mu$ and a possible $7\sigma$ excess in the mass of $W$ boson it is imperative to investigate new physics contributions for clarification. We propose the ScotoZee model, a simple charged singlet extension of the Scotogenic model, to show a direct correlation between these anomalies and the observed neutrino oscillation data as well as dark matter relic abundance. We explore the parameter space spanned by both the bosonic and fermionic dark matter candidates and provide a coherent resolution to electron and muon AMMs and $M_W$ anomaly while evading dangerous LFV processes like $\mu \to e\gamma$ and $\mu \to 3 e$. In contrast to the IDM/Scotogenic models where the small mass splitting among the doublet fields required for the observed relic density is disfavored by the CDF measurement, the scalar DM candidate in our model survives due to the presence of the extra charged singlet which is essential in resolving $\Delta a_\ell$. This model predicts large rates for LFV processes $\tau \to \ell \gamma$ which can be tested in the future experiments.

\textbf{\textit{Acknowledgements:}}
We would like to thank Julian Heeck, K.S. Babu, Vishnu P.K., and D. Raut for useful discussions. RD thanks the U.S.~Department of Energy for the financial support, under grant number DE-SC 0016013.

\section*{Appendix: Oblique parameters}
\label{App-01}

Here we provide additional plots in the model to show the allowed parameter space consistent with the upward CDF W mass shift.

\begin{figure}[!h]
    \includegraphics[width=0.45\textwidth]{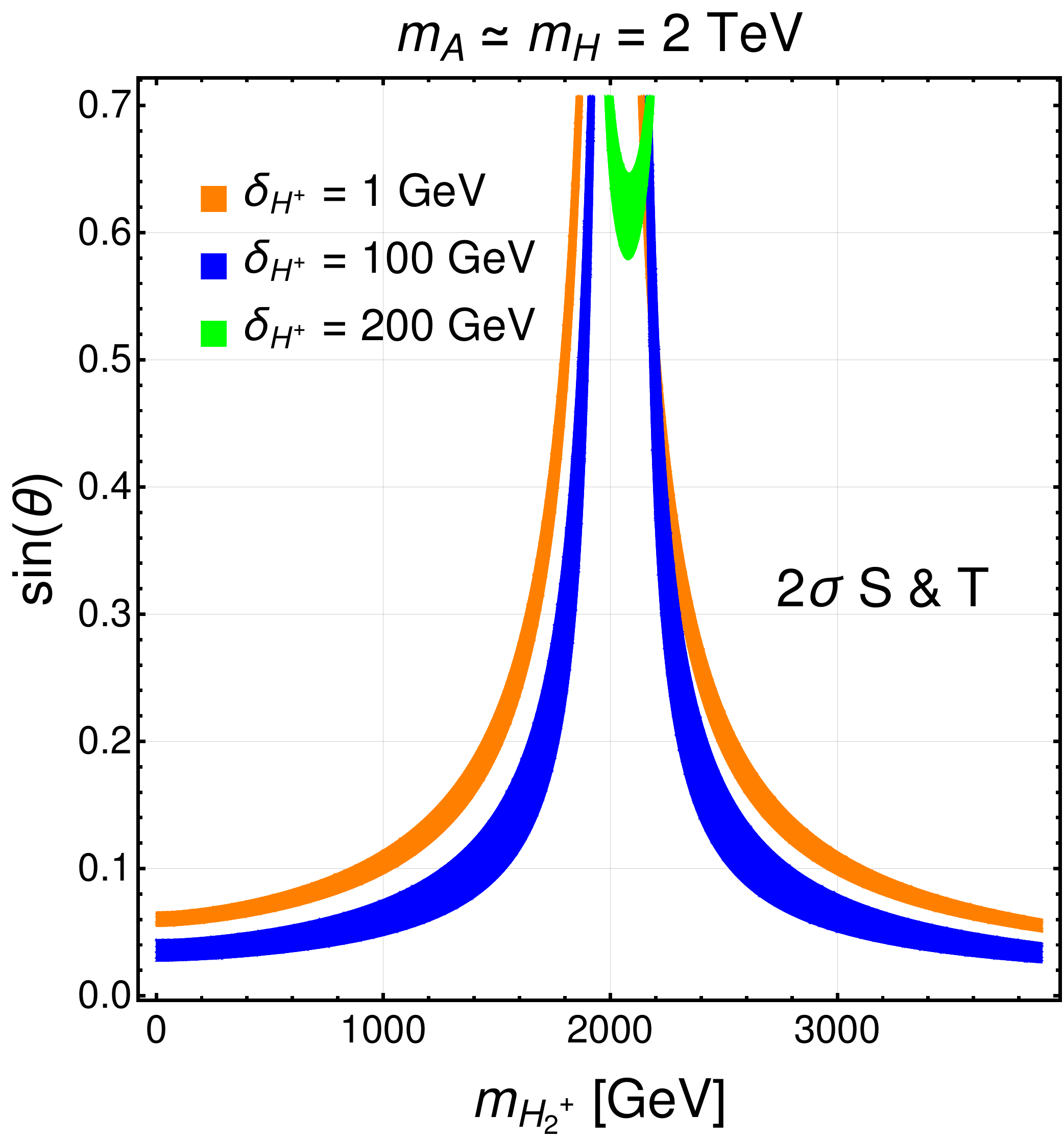} \hspace{10mm}
     \includegraphics[width=0.45\textwidth]{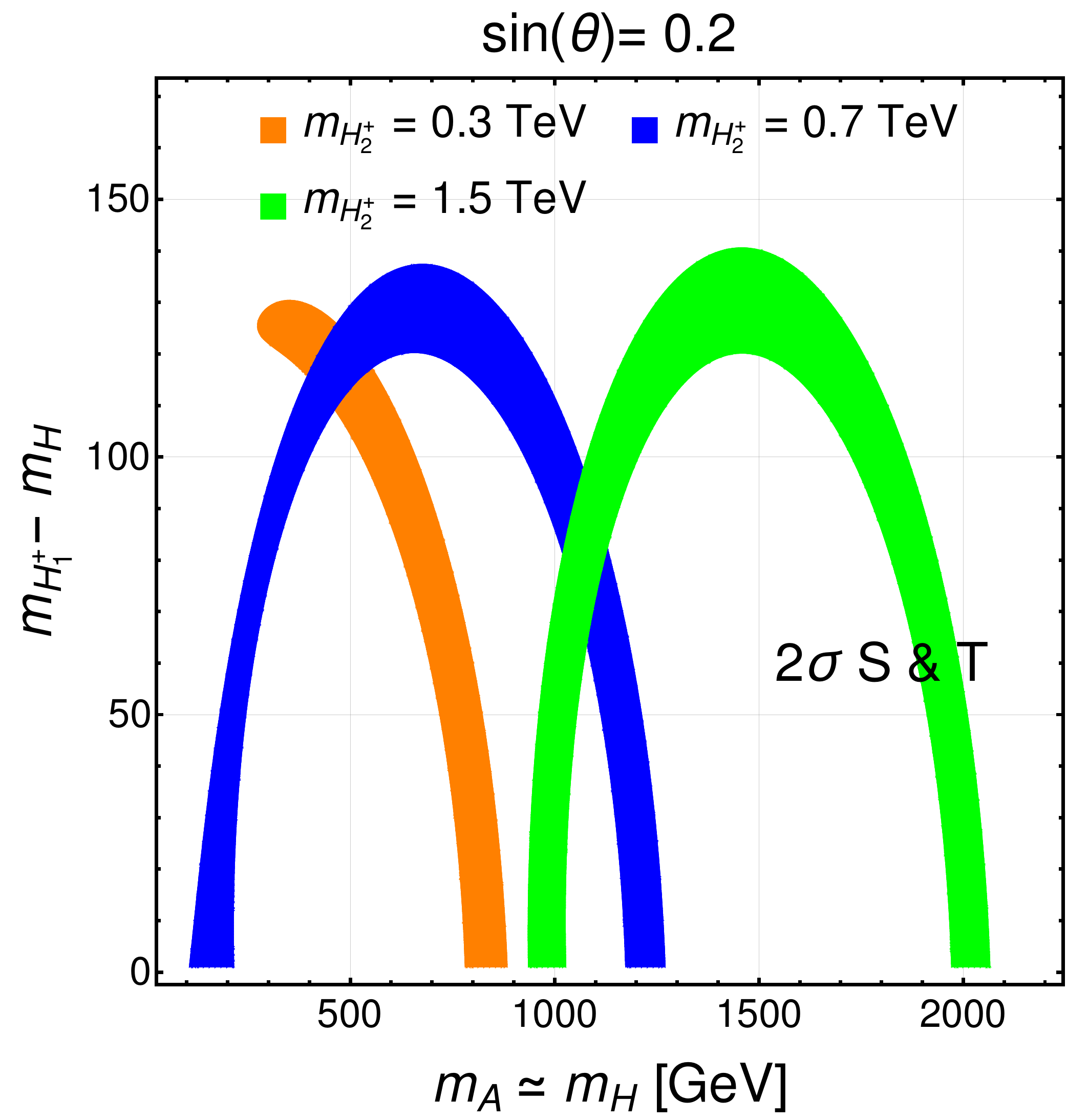}
    \caption{Mixing angle~$\theta$ as a function of charged scalar mass $m_{H_2^+}$ for different mass splitting, $\delta_{H^+}=m_{H_1^+}-m_H$~(left) and the mass splitting $\delta_{H^+}$ as a function of neutral scalars for different choices of charged singlet scalar mass~(right)  explaining the upward shift in $M_W$ reported by CDF measurement, consistent with the $2\sigma$ ranges of $S$ and $T$.}
    \label{fig:Tparam2}
\end{figure}

\bibliographystyle{utcaps_mod}
\bibliography{BIB}

\end{document}